# Wi-Fi Based Passive Human Motion Sensing for In-Home Healthcare Applications

Bo Tan, Alison Burrows, Robert Piechocki, Ian Craddock
Department of Electrical and Electronic Engineering,
University of Bristol, BS8 1UB, UK
{b.tan, alison.burrows, r.j.piechocki, ian.craddock}@bristol.ac.uk

Karl Woodbridge, Kevin Chetty
Department of Electronic and Electrical Engineering,
University College London, WC1E 7JE, UK
{k.woodbridge, k.chetty}@ucl.ac.uk

*Abstract*— This paper introduces a Wi-Fi signal based passive wireless sensing system that has the capability to detect diverse indoor human movements, from whole body motions to limb movements and including breathing movements of the chest. The real time signal processing used for human body motion sensing and software defined radio demo system are described and verified in practical experiments scenarios, which include detection of through-wall human body movement, hand gesture or tremor, and even respiration. The experiment results offer potential for promising healthcare applications using Wi-Fi passive sensing in the home to monitor daily activities, to gather health data and detect emergency situations.

## I. Introduction

The ageing of the population and rise in chronic health conditions have had significant economic and social implications for public sectors, family and individuals. This has precipitated a shift towards understanding daily activities and well-being at home in order to make lifestyle changes to manage and even prevent chronic health conditions. An Internet of Thing (IoT) approach to monitoring people's daily living activities for the purposes of early diagnosis, monitoring chronic disease, and making informed lifestyle changes is proposed in [1]. The SPHERE project aims to collect data on the home environment and residents' activities via a wireless sensor network. Activity data will be collected through wearable accelerometers and video system, with a focus on body pose, motion and activity intensity that may be linked with various health conditions. However, there are concerns that the wearable device may not always be worn by the users and the video system will not cover certain areas such as bedrooms and bathrooms due to the privacy issues. Thus, a Wi-Fi-based passive motion sensing concept is proposed in this paper, with a view to underpinning a non-intrusive and full coverage in-home healthcare IoT platform.

Passive wireless sensing (PWS) has evolved from passive radar, which leverages wireless signals for non-cooperative detection. Research has focused on using a wide range of transmission sources including GSM, FM, WiMax [2] and Wi-Fi [3], to detect targets such as aircrafts, ships, vehicles and humans. The use of 802.11 signal as an illuminator for target discovery has been reported in [4] for outdoor vehicle tracking, and [3] for indoor through-wall human discovery. Subsequent works on 802.11-based passive radar are further extended to real time [5] and small human body movement [6] detection capability. Most passive or active small movement detection systems use micro Doppler as the main metric. The detection and utilization of micro Doppler has been reported in [7] by using ultrasound, and [8] by using 800 MHz bandwidth microwave signal. To use of 2.4 GHz ISM band wireless signal for human movement and motion capture is also reported in [9, 10]. In [9], a Doppler shifting extracting method for OFDM signal is described and verified with specially generated OFDM signal source. In order to achieve the Doppler shift resolution, the reflected OFDM symbols are demodulated and decoded by following standard 802.11 protocol, then equalized with the first received OFDM symbol. To implement the method, the hardware or software decoding module is inevitable, and potentially adds to the complexity of the system. From the results shown in [9], there are still strong zero Doppler residue that probably comes from direct signal leak, which is common in passive detection applications. In [10], an angle estimation method is described to detect human movement in indoor scenarios with specially generated OFDM signals on 2.4G ISM band. The performance of this method may be difficult to predict when using commercial off-the-shelf (COTS) Wi-Fi devices, which transmit variable temporal bursts with random interval. To achieve more detailed human activity information, the authors of [10] demonstrated a tracking function with 1.65 GHz ultra-wideband signal FMCW chirp on 6 o 7 GHz carrier frequency in [11]. In [12], the same authors extend the detection ability from tracking to respiration with specific layout of the active ultra-wideband FMCW angle detection system.

This paper proposes a real time 802.11 signal based passive human movement capturing system, which uses Doppler as the main metric. The method used for Doppler extraction in this work is cross ambiguity function (CAF). Due to the low range resolution of 802.11 signals, the echo delay in CAF is ignored in this work. Leveraging fast CAF processing, the proposed system gains the capability of sensing very small and slow human movements by taking longer integration duration. In addition, it is easy to add pre and post processing, for example, normalized least mean square (NLMS) filter or CLEAN algorithm to mitigate the direct signal interference (DSI) to improve the detection performance. The concept is verified with software defined radio (SDR) in different scenarios, which include through-wall human body motion sensing by using the COTS Wi-Fi AP, hand movement and hand tremor sensing by using laptop Wi-Fi signal leak, and even respiration by using 2.4 GHz ISM band CW signal which can be easily transplanted to Wi-Fi signal.

Comparing with works in [9, 10, 11, 12], the proposed Wi-Fi passive sensing technology presents the following advantages:

- The proposed signal processing is a general method that can be applied for all the possible signals, so, the system will not be constrained by OFDM modulated signal as used in [9] and [10].
- The proposed system uses passive detection approach which means that it takes the ubiquitous existing signals in the space to achieve diverse motion detection abilities. The requirement of the high-spec ultra-wideband signal as used in [11] and [12] is not required. In this paper, mostly a non-cooperative 11 MHz bandwidth Wi-Fi signal is used to verify the concept in this paper.
- The proposed method for detecting different types of movements is a signal processing method. In other words, it is a software behaviour. By swapping the software parameters set, different motions can be detected. The hardware embodiment can be modified according to the requirements of each practical scenario.
- The proposed method has been verified with temporal bursts from COTS 802.11 AP. Other works, such as [9] and [10], used the controllable self-generated signals for experiments while the performance may deteriorate if moved to the practical random burst situation.

The rest part of this paper is organized as follows: the next section describes the principle of signal processing and main technology innovations of the PWS system. Section III presents the application scenarios, and the corresponding experimental results are shown and analysed. Conclusions, potential applications, extended functions and future work are given in the last section.

## II. PASSIVE WIRELESS SENSING

### A. Passive Wireless Sensing System

Passive sensing leverages existing transmissions to detect and monitor non-cooperative targets. The illuminator could be, for example, an FM mobile base station, or a DVB or DAB transmitter. In this paper, the illuminator is a wireless network transmitter based on the IEEE standard 802.11, commonly referred as Wi-Fi. Such transmitters are widely deployed in residential and care homes, and are thus ideal for healthcare applications. In order to sense the target, a passive sensing system is equipped with two antennas. One is used for obtaining the signal directly from the illuminator, while the other is used for capturing the reflected signal from the target. The signals from two antennas are called the reference signal and surveillance signal, respectively. The two signals are correlated and processed using the CAF in equation (1).

$$CAF(\tau_d, f_d) = \sum_{n=0}^{N} r[n]s^*[n+\tau_d]e^{-j2\pi f_d \frac{n}{N}} \quad (1)$$

where N is the total number of samples, n means the n-*th* sample, $r[n]$ and $s[n]$ are discrete-time reference and surveillance signals in complex form, operator * is the complex conjugate of the signal, $\tau_d$ is the time delay which can be converted to range information, and $f_d$ is the frequency shift. Using equation (1), it is easy to identify the relative range and velocity of the target in relation to the illuminating transmitter. Relative range is indicated by the time delay $\tau_d$ of the surveillance signal. Relative velocity is indicated by the frequency shift $f_d$ of frequency shift. When there is some finite movement of the target, a clear signal should be present for the corresponding values of $\tau_d$ and $f_d$ on the CAF surface.

Parameters $\tau_d$ and $f_d$ in equation (1) quantify the effective range and frequency resolution of the passive radar systems in [13]. In the case of passive Wi-Fi radar, the range resolution is limited by the bandwidth of the 802.11 signal, which results in 10 to 15 meter range resolution. This is not accurate enough for most indoor applications apart from simple motion sensing. The frequency resolution in equation (1) is determined by the integration time, which is equivalent to the number of samples in the discrete system. Compared with the signal bandwidth, the number of samples is a controllable parameter in the PWS system. By increasing the number of samples taken in equation (1), the system sensitivity to velocity can be improved to a high level. Some very small and slow movements can therefore be captured with the system. The main drawback of the long integration time is the increase in processing time of the data. This problem has been resolved with by the CAF calculation methods introduced in this paper. This will be described in this section, along with the interference cancellation and pipeline processing flow design.

A basic passive wireless system has three main components: antennas for detecting the reference and surveillance channel signals, ADC module for each channel and a computing unit for signal processing. The specification of each component varies according to the application scenario. The antenna can range from high gain, directional dish types to a miniature low gain patch. The ADC module could be, for example, a general purpose USRP module or use a dedicated ADC front end for 2.4 GHz signals. The computing unit could be a PC, an embedded system, or DSP chips. The demonstration PWS system used in this paper is described in section IV.

### B. Real Time Signal processing

As mentioned in the previous subsection, longer integration times lead to higher Doppler sensitivity, while resulting in longer processing times and increased computing load. This makes it very difficult to carry out real time processing without avoiding information drop out, if simply using the calculation method in equation (1). In order to enable the real time sensing that is required in most healthcare scenarios, three techniques are implemented in the PWS system: pipeline processing, down sampling and batch processing.

*Pipeline:* Pipeline processing is an extension of parallel code execution that enables performance gains with serial multistage algorithms on multicore machines. A sequential code sequence is partitioned into sub-procedures, which are each allocated to a separate core. The algorithm can then run simultaneously on multiple sets of recorded data, or data that streams continuously. To maximize throughput, each subroutine should be carefully balanced to ensure approximately equal processing times. The delay of a pipelined system can be written as (2):

$$\tau_p = max\{\tau_i\} \qquad (2)$$

where $\tau_i$ is the delay of $i$-th sub-procedure. In the PWS system, the sub-procedures are: signal sampling, CAF processing, interference cancellation, which are shown in Fig. 1.

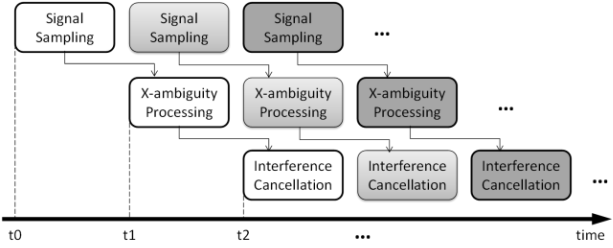

Fig. 1. Pipeline processing architecture in wireless passive radar

***Down sampling:*** As previously mentioned, range resolution is not an issue in the Doppler sensing mode used in the healthcare context. The sampling rate can therefore be reduced in order to achieve higher Doppler sensitivity by taking longer integration time with the same number of samples. The aim of the PWS is not to decode the information, thus the down sampling will not impact the Doppler detection.

***Batch Processing:*** Batch processing is a concept taken from IT systems, which is used for executing a series of non-interactive tasks. In PWS context, the pending samples are divided into segments and then only the beginning of the each segment is taken for processing. The processing principle can be found in [14]. The processing flow of batch processing is shown in Fig. 2.

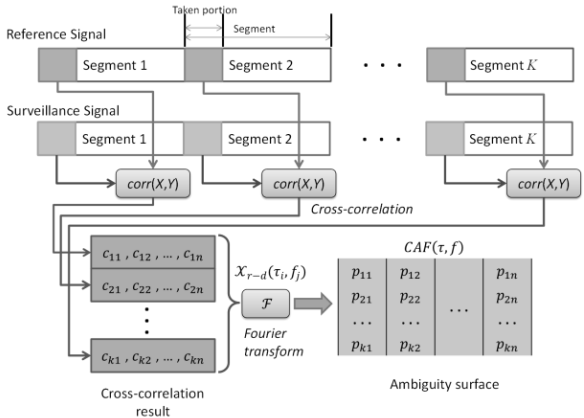

Fig. 2. Batch processing flowchart

The recorded synchronized reference signal and surveillance signal are divided into $v$ isometric data segments respectively. For each segment, only a certain portion of the sampled data are taken and used for the follow-up processing. Experimental results show that a 10% portion of the signal is sufficient to build a clear ambiguity surface. The number of segments $J$ is determined by the predicted maximum target velocity $v_{max}$. The relationship can be described by the following equation:

$$J = 2 \cdot \left\lceil \frac{v_{max}}{c} \cdot f_0 \cdot \tau_{sample} \right\rceil \qquad (3)$$

where operator $\lceil x \rceil$ denotes rounding the elements of $x$ to the nearest integers towards infinity, $c$ is the propagation velocity of the wireless signal, and $f_0$ is the centre frequency. The factor 2 is used for reserving $v_{max}$ for both forward and backward directions.

Besides the above processing, other processing techniques such as CLEAN [4] and CFAR [15] are also applied after the CAF processing for interference suppression and target detection. The CLEAN algorithm is applied here to cancel the interference from direct signal (DSI) and the reflections from the stationary objects, by subtracting a scaled reference surface which is constructed by reference signal. The scaling factor is updated in each iteration [4]. CFAR is widely used in radar detection systems for detecting the target by setting a threshold of the index which can be derived from different criteria; this is beyond the scope of this paper. The criterion in this paper is called cell average [15].

### III. EXPERIMENT SCENARIOS AND RESULTS

In this section, results from three different indoor sensing scenarios are presented to demonstrate the how the signal processing described in the previous section is applied in a real environment. The application scenarios include: through-wall body gesture detection, hand gesture/tremor sensing, and breathing detection. The experimental setup and detection results are also provided. The PWS system used in the following experiments is a SDR based system using USRP modules. A schematic of this system is shown in Fig. 3. In this system two synchronized Ettus USRP N210s are used for downconverting the RF signal and digitizing. The laptop is equipped with an Intel Core i7 3.2 GHz CPU. The sampled data is imported to the laptop via Ethernet ports. The operating system provides multiple thread options for implementing pipeline processing in a Labview software.

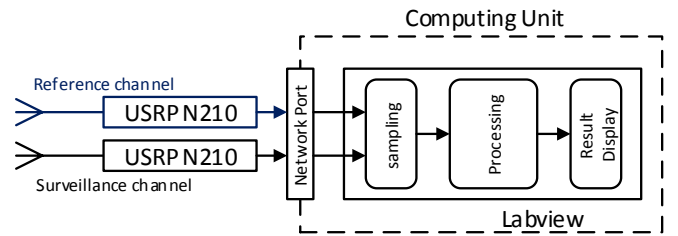

Fig. 3. Block diagram of the real-time software-defined wireless passive

#### A. Through-Wall Body Gesture Detection

In this scenario, the main aim is to detect human body movement through-wall or non through-wall for in-home healthcare applications. In this section the SDR based system (Fig. 3) is set up for human body gesture detection through a 22cm brick wall in Fig. 4. The Wi-Fi AP used in this experiment is an Edimax M3000, which has 15 dBm power emissions. One high gain antenna points to the AP as the reference channel, and the other antenna is directed to the target as the surveillance channel. Four different body gesture examples and corresponding Doppler records are shown in Fig. 5.

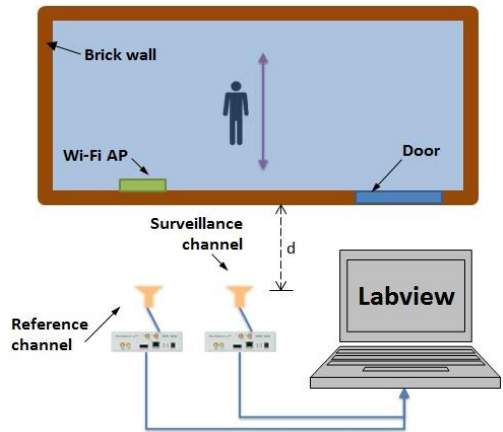

Fig. 4. The experimental setup for through-wall human body gesture detection

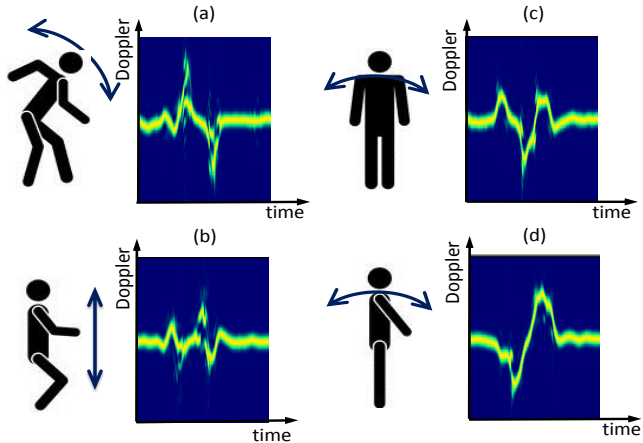

Fig. 5. Human body movement detection results: (a) stooping-back, (b) squating-standing, (c) left-right body swing, (d) forward-backward swing. A 4 seconds Doppler record is shown for each guesture cycle. The dynamic Doppler shift range of the body gesture is around ±10 Hz

From the experiment results, we can see that different body gestures cause distinctive Doppler records. So, in-home human activities could be recognized using a template library and suitable classifiers. The recognized information is valuable for diverse health-related applications. For example, in [16] the authors have clearly identified links between the intensity of daily activity and certain psychological conditions. Also, the detection of specific activities can be used for alerting carers to emergency situations such as falling or lack of movement for long periods.

*B. Hand movement and Tremor Detection*

Another potential healthcare application for the system is in the detection of small movements such as hand tremors. In this case, the system takes advantage of the embedded Wi-Fi emission from mobile devices, such as laptops and tablets, to form the wireless passive radar. The setup for this experiment is shown in Fig 6.

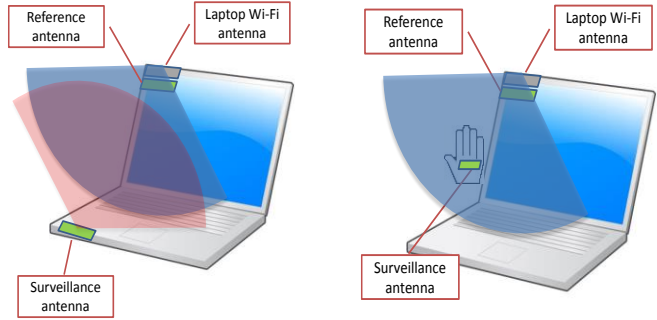

Fig. 6. Layout of the hand movement detection experiment on a laptop. Left: the hand-free setup, Right: attached solution.

Very small PCB antennas are used for receiving the reference and surveillance signal. The surveillance antenna can be fixed in position on the laptop, or attached to the hand for fine movement measurements. Fig. 7 shows results for four different movements with the PCB antenna mounted on the laptop. Fig. 8 shows detection of very small tremors with the PCB antenna attached to the hand. These tremors are typical of those found in neurological disease such as Parkinson's.

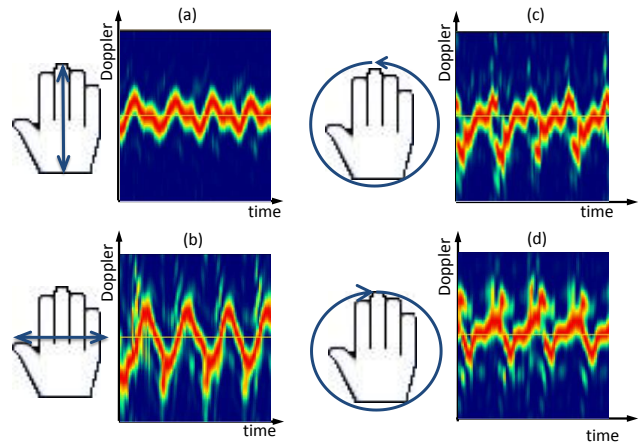

Fig. 7. Hand movement detection results. (a), forward-backward movement, (b), left to right movement, (c), counterclockwise movement, (d), clockwise movement. 4 second Doppler record is shown for each gesture

In Fig.7 and Fig. 8, 4 seconds Doppler records are shown to demonstrate the capability of the system for hand motion and hand tremor detection. Fig.7 shows distinguishable Doppler pattern for each hand motion. In Fig. 8, the fast tremor causes the higher frequency jitter, which is about 4~6 Hz, and the slower tremor causes the lower frequency jitter, which is around 2~3 Hz. The measured tremor frequency could be of value to clinicians to distinguish tremor due to different neurological conditions. A wearable or device free system could also be implemented enabling collection of tremor data over extended periods. This would be very valuable for monitoring the long-term evolution of the condition.

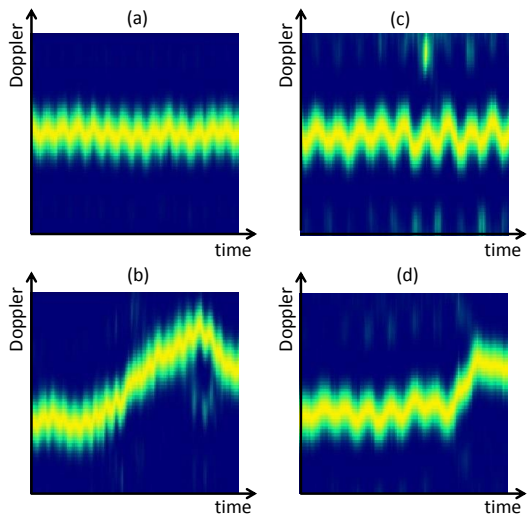

Fig. 8. Hand tremor detection results: (a) pure hand tremor with higher tremor frequency, (b) combined hand movement and tremor at higher frequency, (c) pure hand tremor with lower tremor frequency, (d) combined hand movement and tremor at lower frequency.

*C. Breathing Detection*

Another important application of this technology is for detecting signs of life. In this section, an experiment is described that uses the PWS principle to detect bio-activity, specifically chest wall movement associated with breathing. The antennas and layout of the experiment is similar to that shown in Fig. 4, without the intervening wall. In this preliminary experiment, the Wi-Fi illuminator is replaced by a 2.4GHz Continuous-wave (CW) signal source with an output power 10 dBm lower than the Edimax 3000 Wi-Fi AP used in the previous experiment. The experimental data is processed offline with Matlab.

In contrast to the post interference cancellation method used in the previous scenarios, pre interference cancellation is applied to mitigate the direct signal leakage from the signal source. The pre interference cancellation method used in this case is the NLMS. Fig. 9 shows results from a human subject, recording about 7 respirations over about 25 seconds. On the Doppler record surface generated with the filtered surveillance signal, the positive and negative Doppler shifts caused by the chest rising and falling movements can clearly be seen. The breathing frequency detected is close to the typical adult human respiration rate of 12~15 per minute. This method used for the CW transmitter can be extended to down sampled bandwidth Wi-Fi signals by properly adjusting the step size in the NLMS filter for the breathing movements. This preliminary demonstration of passive breathing detection shows that the PWS could be a valuable tool in monitoring signs of life in hospitals or residential homes.

IV. CONCLUSION AND FUTURE WORK

This paper described the signal processing method, system combination and experiment scenarios. The experimental results show that Wi-Fi based passive sensing technology is suitable for non-intrusive data collection of human movement and activities, ranging from gross body movements to small hand tremors. It is envisaged that this technology can be implemented in various scenarios and for diverse purposes.

Wi-Fi APs are widespread in residential area, thus providing ample opportunities for PWS applications. The through-wall detection ability of this technology lends itself to broad application scenarios especially in-home healthcare, for example to unobtrusively detect falls, abnormal events, and daily activities. This paper details the use of PWS to detect hand tremor and respiration, which have the potential to be used to further understand a range of health conditions as well as to detect extreme events. This technology also has the potential to enable alternative forms of interaction within the smart home such as using movement to control settings or using breathing rates to respond to mood, without the need for dedicated and sometimes intrusive devices.

Although the technology is still in its infancy, it has the potential to be further developed according to different directions. The first direction is to applying machine learning technology to interpret the premier Doppler detection results. Using machine learning methods, more detailed instantaneous human body and gesture information can be sensed, while the long term passive sensing data can be used to understand the relation between daily routines and health conditions. Secondly, it can be integrated with data from different types of sensors within the IoT. As there are also video, accelerometers and environmental sensors on the SPHERE platform [1], fusing these data with passively detected Doppler information will add a new dimension to discovering unique features of daily movements. Lastly, the signal processing of passive sensing technology is far from completion; there is still large area to explore new signal processing methods and functions. For example, by synthesising the Doppler detections from different locations, high accuracy target moving trace can be calculated with proper tracking filter calibration. High accuracy location information is a key requirement in many IoT applications. The idea has been tested and summarized in our recent paper [17], which constrains the localisation error down to half meter for the indoor scenario.

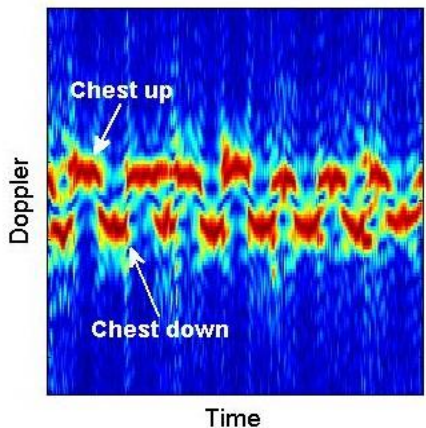

Fig. 9. Detection result for human breathing using a 2.4 GHz CW transmission